\documentclass[twocolumn,preprintnumbers]{revtex4}
\usepackage{amssymb}

\usepackage{graphicx}
\usepackage{dcolumn}
\usepackage{bm}

\begin{document}

\newcommand{\li}{$^6$Li}
\newcommand{\lit}{$^6$Li$_2$}
\newcommand{\na}{$^{23}$Na}
\newcommand{\cs}{$^{133}$Cs}
\newcommand{\kk}{$^{40}$K}
\newcommand{\rb}{$^{87}$Rb}
\newcommand{\vect}[1]{\mathbf #1}
\newcommand{\g}{g^{(2)}}
\newcommand{\one}{$\left|\uparrow\right>$}
\newcommand{\two}{$\left|\downarrow\right>$}
\newcommand{\V}{V_{12}}
\newcommand{\kfa}{\frac{1}{k_F a}}
\newcommand{\hbk}{$\hbar k_L$}
\newcommand{\Er}{$E_r$}

\title{Evidence for Superfluidity of Ultracold Fermions in an Optical Lattice}

\begin{abstract}
The study of superfluid fermion pairs in a periodic potential has
important ramifications for understanding superconductivity in
crystalline materials. Using cold atomic gases, various condensed
matter models can be studied in a highly controllable environment.
Weakly repulsive fermions in an optical lattice could undergo
d-wave pairing \cite{hofs02hightc} at low temperatures, a possible
mechanism for high temperature superconductivity in the cuprates
\cite{scal95hightc}. The lattice potential could also strongly
increase the critical temperature for $s$-wave superfluidity.
Recent experimental advances in the bulk include the observation
of fermion pair condensates and high-temperature superfluidity
\cite{grei03mol_bec, joch03bec, zwie03molBEC, rega04,
zwie04rescond, zwie05Vort}.  Experiments with fermions
\cite{modu03ferm, kohl05ferm, stof06} and bosonic bound pairs
\cite{volz06, wink06} in optical lattices have been reported, but
have not yet addressed superfluid behavior. Here we show that when
a condensate of fermionic atom pairs was released from an optical
lattice, distinct interference peaks appear, implying long range
order, a property of a superfluid. Conceptually, this implies that
strong $s$-wave pairing and superfluidity have now been
established in a lattice potential, where the transport of atoms
occurs by quantum mechanical tunneling and not by simple
propagation.  These observations were made for unitarity limited
interactions on both sides of a Feshbach resonance.  For larger
lattice depths, the coherence was lost in a reversible manner,
possibly due to a superfluid to insulator transition.  Such
strongly interacting fermions in an optical lattice can be used to
study a new class of Hamiltonians with interband and atom-molecule
couplings \cite{duan05}.
\end{abstract}

\author{J. K. Chin, D. E. Miller, Y. Liu, C. Stan, W. Setiawan, C. Sanner, K. Xu, W. Ketterle}

\homepage[Group website: ]{http://cua.mit.edu/ketterle_group/}
\affiliation{Department of Physics, MIT-Harvard Center for
Ultracold Atoms, and Research Laboratory of Electronics,
Massachusetts Institute of Technology, Cambridge, MA 02139}
\maketitle

%\draft               % preprint mode

%\newcommand{\lambdabar}{ \mathchar'26\mkern-10mu\lambda}

Previous experiments demonstrating long-range phase coherence in
Bose-Einstein condensates (BECs) and in fermion superfluids used
ballistic expansion to observe interference of two independent
condensates \cite{andr97int}, vortex lattices
\cite{zwie05Vort,madi00, abos01latt} or interference peaks after
release from an optical lattice \cite{ande98atla, grei01mott}.
However, for strongly interacting fermions, elastic collisions can
change the momentum distribution and wash out interference peaks.
For an initially superfluid cloud, such dissipative dynamics
corresponds to superfluid flow faster than the critical velocity.
Consistent with this expectation, expansion of the strongly
interacting Fermi gas from an optical lattice yielded a diffuse
cloud which exhibited no signs of interference (Fig. 1).  This issue
was addressed by using a magnetic field ramp which quickly increased
the detuning from a Feshbach resonance, taking the system out of the
strongly interacting regime and enforcing ballistic expansion. In
previous studies of strongly interacting Fermi gases magnetic field
sweeps were applied to prevent fermion pairs above the Feshbach
resonance from dissociating
\cite{rega04,zwie04rescond,schu06expand}. In contrast, our
experiment required a magnetic field sweep both above and below the
Feshbach resonance to avoid elastic collisions.

\begin{figure}
    \begin{center}
    \includegraphics[width=3.3in]{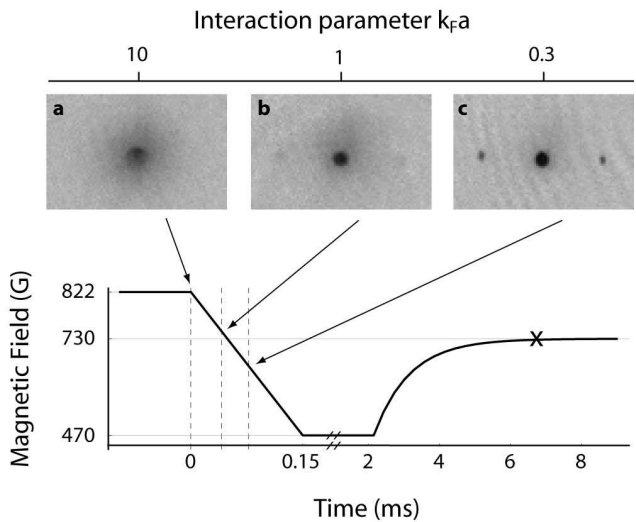}
    \caption[Dissipative collisions during expansion of a strongly interacting fermionic superfluid.]{
    The schematic shows the  time sequence of the magnetic
    field ramp used throughout this Letter.  A
    one-dimensional optical standing wave was pulsed onto the superfluid
    at different values $B_p$ (indicated by arrows (\textbf{a}) 822 G,
    (\textbf{b}) 749 G and (\textbf{c}) 665 G) of the magnetic
    field during expansion, creating particles at twice the photon recoil \cite{goul86}.
    Absorption images taken at the time marked with the cross show distinct
    momentum peaks only at magnetic fields $B_p \leq 750 G$,
    corresponding to $k_F a \leq 1$ .  At higher magnetic fields,
    the peaks blurred into a broad diffuse cloud as a result of the larger
    collision cross-section. } \label{fig:kapitza}
    \end{center}
\end{figure}

Our experiments used a balanced mixture of \li  fermions in the
two lowest hyperfine states.  Evaporative cooling produced a
nearly pure fermion pair condensate which was adiabatically loaded
into a three-dimensional optical lattice. A broad Feshbach
resonance centered at 834 G enabled tuning of the interatomic
interactions over a wide range. On resonance, a bound molecular
state becomes degenerate with the open atomic scattering channel,
leading to a divergence in the scattering length $a$.  Here we
explore the region of strong interactions, also known as the
BEC-BCS (Bardeen-Cooper-Schrieffer) crossover, where the magnitude
of the interaction parameter $|k_F a|$ is greater than unity, and
$k_F$ is defined as the peak Fermi wavevector of a two-component
non-interacting mixture of \li\ atoms.  Throughout this regime,
pairing occurs as a result of many-body interactions. Below
resonance, for strong interactions, the bare two-body state has a
bond length larger than the interatomic spacing and is irrelevant.
In a lattice, atom pairs above and below the resonance can be
confined to one lattice site \cite{stof06}, and crossover physics
may require an occupation larger than or equal to one.

The peak pair filling factor of the lattice was about unity.  At
this density in the bulk, the fermion pair size is on the order of
$1/k_F = 170 $ nm, comparable to the lattice spacing of 532 nm. To
probe the momentum distribution, we ramped the magnetic field out
of the strongly interacting regime as fast as technically possible
($\sim$ 150 $\mu$s) and then turned off the confining potential.
Absorption images taken after 6.5 ms of expansion reveal sharp
peaks at the reciprocal lattice vectors --- the signature of
long-range coherence and superfluidity.

\begin{figure}
    \begin{center}
    \includegraphics[width=3.3in]{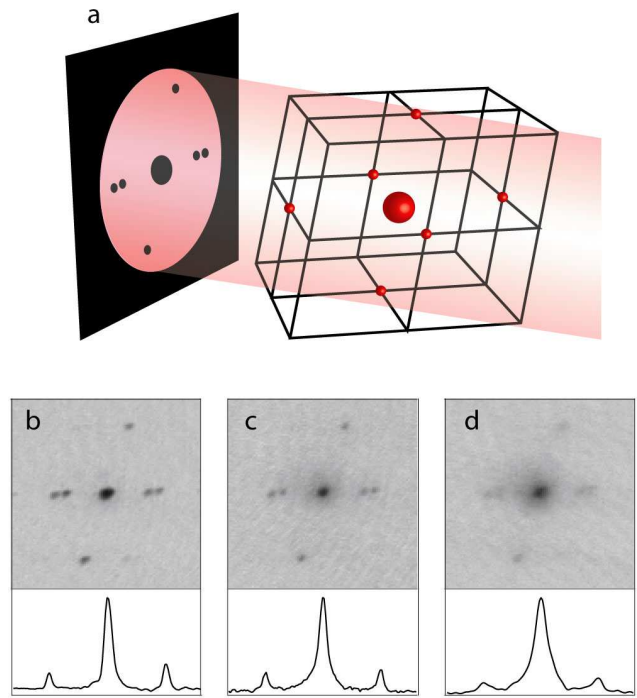}
    \caption[Observation of high contrast interference of fermion
    pairs released from an optical lattice below and above the Feshbach
    resonance.]{(\textbf{a}), The orientation of the reciprocal
    lattice, also with respect to the imaging light. \textbf{b-d}, Interference peaks are
    observed for magnetic fields of (\textbf{b}) 822 G, (\textbf{c}) 867 G
    and (\textbf{d}) 917 G. The lattice depth for all images is 5 \Er\, and each
    image is the average of 3 shots. The field of view is 1 mm by 1 mm. Density profiles through
    the vertical interference peaks are shown for each image.} \label{fig:mainpic}
    \end{center}
\end{figure}

We observed such interference peaks at magnetic fields both above
and below the Feshbach resonance (Fig. 2).  The six first order
diffracted peaks are clearly visible around the zero momentum
fraction and their positions correspond to the expected momentum
quanta of 2\hbk\ carried by molecules of mass 2$m$, where $k_L$ is
the lattice wavevector.  At high magnetic fields (Fig. 2d) the
visibility of the interference peaks decreased and some additional
heating was observed. This degradation could be due to a higher
fraction of thermal atoms as we approached the BCS limit, but was
not studied in detail.

The narrow interference peaks clearly reveal the presence of a
macroscopic wavefunction possessing long range phase coherence.
The separation between the interference peaks relative to their
width gives an estimate of the coherence length of $\sim 10$
lattice sites. This estimate is a lower bound, because effects of
finite resolution and mechanisms of residual broadening have been
neglected. With unity occupation, and in the absence of any
discernible background at magnetic fields near the Feshbach
resonance, this implies a minimum phase space density of $10^3$,
and shows that our samples are deep in the quantum-degenerate
regime. In previous studies of ultracold Bose and Fermi gases, the
appearance of a condensate fraction and long range phase coherence
was shown to occur concurrently with the possibility to excite
superfluid flow \cite{zwie05Vort,madi00,abos01latt,onof00sf}.
Superfluid hydrodynamics is usually regarded as the direct proof
for superfluidity. However, all reports of superfluidity of bosons
in three-dimensional optical lattices have relied solely on
observations of sharp interference peaks and inferred
superfluidity from the established connection between long-range
coherence and superfluidity \cite{grei01mott,scho04exlat}.
Similarly, our observations directly show long-range coherence and
indirectly show superfluidity of fermion pairs in an optical
lattice.

For deep lattices, breakdown of superfluid behavior has been
observed \cite{grei01mott, xu05} for weakly interacting
Bose-Einstein condensates. This phase transition to the
Mott-insulator state occurs when on-site interactions start to
suppress atom number fluctuations and the system transitions from
a delocalized superfluid described by a macroscopic wavefunction,
to a product of Wannier states tightly localized at each lattice
site. Experimentally, this is manifested as a smearing of the
distinct 2\hbk\ interference peaks.

Fig. 3 shows the evolution of a strongly interacting fermionic
superfluid when the lattice depth was increased. The interference
peaks became more pronounced initially, due to increased
modulation of the wavefunction. The interference peaks began to
smear out, rapidly giving way to a featureless cloud, beyond a
critical lattice depth $V_{c}\approx 6$ \Er\, where $E_r=\hbar^2
k_L^2/4m = h\times$ 15 kHz is the recoil energy. This indicates
that all phase coherence had been lost. Upon subsequent ramp down
of the lattice, interference peaks became visible again (Fig. 3h),
demonstrating reversibility of the lattice ramp.

\begin{figure}
    \begin{center}
    \includegraphics[width=3.3in]{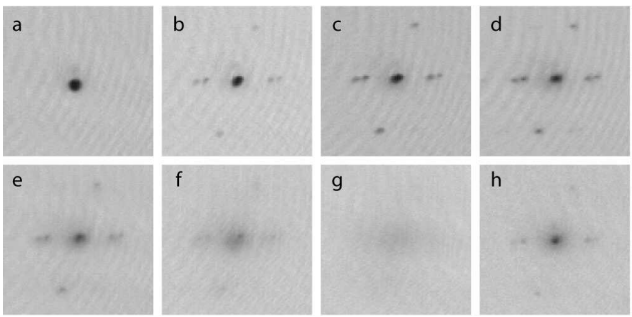}
    \caption[Interferograms of fermion pairs released from different lattice
    depths $V_0$ at a field of 822 G.]{Values of $V_0$'s are (\textbf{a}) 0 \Er\ , (\textbf{b}) 2.5 \Er\ ,
    (\textbf{c}) 4 \Er\ , (\textbf{d}) 5 \Er\ , (\textbf{e}) 6 \Er\ , (\textbf{f}) 7 \Er\ ,
    (\textbf{g}) 9 \Er\ , (\textbf{h}) 2.5 \Er\ .
    (\textbf{a}-\textbf{g})
    were taken after an adiabatic ramp up to the final
    $V_0$, while (\textbf{h}) was taken after first ramping up to 10 \Er\ ,
    before ramping down to 2.5 \Er\ .} \label{fig:rampup}
    \end{center}
\end{figure}

We repeated this sequence for a wide range of initial magnetic
fields both above and below the resonance and observed the same
marked change in the interference pattern. Fig. 4 displays the
peak optical density of the interference peaks for different
lattice depths at representative fields. Across all fields, the
sharp decrease in peak optical density occurred between 5 and 6
\Er\ . Further increase of the magnetic field resulted in
decreasing overall visibility, until interference peaks could no
longer be observed regardless of lattice depth.

\begin{figure}
    \begin{center}
    \includegraphics[width=3.3in]{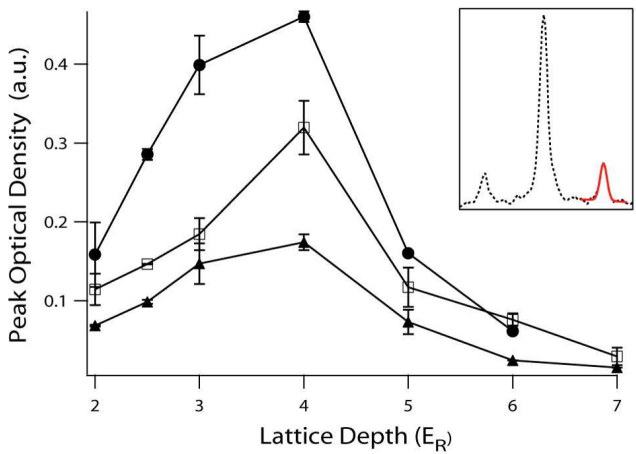}
    \caption[Peak optical density of interference peaks for increasing lattice
    depths at different magnetic fields.]{
    Values of magnetic fields are 842 G (filled circles), 892 G
    (open squares) and 942 G (filled triangles). Peak optical densities were estimated
    from fits to the peaks, including background subtraction. The inset (color online) shows
    a sample density profile of the central and one pair of interference peaks (black dotted
    line), with a bimodal fit to one side peak (red solid line). Each point is the
    average of three different images with six interference peaks per image.
    Error bars show s. d.}
\label{fig:widthfield}
    \end{center}
\end{figure}

The loss of phase coherence with increasing lattice depth is
consistent with the qualitative description of the superfluid to
Mott-insulator transition. However, the usual single-band
description is no longer applicable, since in the strong-coupling
regime the on-site interaction strength should be comparable to
the band gap $\hbar \omega$, where $\omega$ is the onsite trap
frequency. Furthermore, Pauli blocking  forbids the multiple
occupation of the lowest state of an individual lattice site by
identical fermions, and modification of the single particle
tunneling rate is expected due to virtual pair breaking
transitions \cite{duan05}. One may still be tempted to use the
standard bosonic Hubbard model and estimate the critical lattice
depth $V_c$ for an assumed value of onsite interaction energy
$U=\hbar\omega$ and non-interacting, single particle tunneling
$J$, but the obtained $V_c \approx 3$ \Er\ is significantly
smaller than our observation, which is in turn much smaller than
the $V_c
> 10$ \Er\ observed for weakly interacting atomic BECs
\cite{grei01mott, xu05}.  Along with the observed insensitivity of
$V_{c}$ to magnetic field, this demonstrates that models based on
weak interactions are inadequate.

Fig. 3h demonstrates the reversibility of the transition from a
long range coherent state to a state without strong coherence. We
now study the time scale for this recoherence, in analogy with
similar measurements performed across the superfluid to
Mott-insulator transition in atomic BECs \cite{grei01mott}. Fig. 5
shows that phase coherence was restored on a sub-millisecond time
scale, on the order of the the single-particle tunneling time of
about 500 $\mu$s (for a shallow lattice of 2.5 \Er\ ).  When the
same lattice ramp sequence was applied to a superfluid which had
been dephased by a magnetic field gradient \cite{grei01mott} the
system did not regain phase coherence on the time scales that we
probed. Therefore, evaporative cooling is negligible during this
time.  The short recoherence time of the condensate is evidence
that the system stayed in its ground state or at least in a low
entropy state when the lattice was ramped up.

\begin{figure}
    \begin{center}
    \includegraphics[width=2.5in]{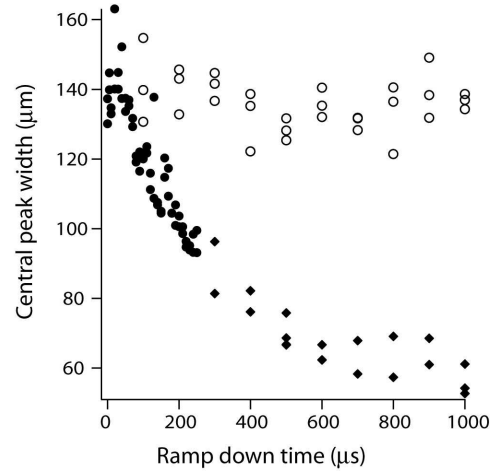}
    \caption[Restoring coherence from a deep lattice. ]{The
    width of the central peak is used as a measure of phase
    coherence after an adiabatic ramp up to 8 \Er\, followed by a fast ramp
    down to 2.5 \Er\ at a fixed magnetic field of 822 G. Filled circles were
    extracted with the use of a gaussian fit, and diamonds with a bimodal fit.
    Also plotted for comparison is the gaussian
    width of the central peak for a dephased sample, in which a field gradient was applied during the ramp
    up of the lattice (open circles). All points were taken for 6.5 ms time-of-flight.}
\label{fig:recoherence}
    \end{center}
\end{figure}

Fig. 5 also provides evidence that the system could not recohere
during the 150 $\mu$s magnetic field ramp. In Fig. 3h, the central
peak is well fitted by a bimodal distribution with a width of 35
$\mu$m, in clear contrast to the gaussian width of 105 $\mu$m
obtained from Fig. 5 after 150 $\mu$s. Therefore, we conclude that
the observed interference patterns in Fig. 1 reflect the coherence
of the cloud at the initial magnetic field, in the strongly
interacting regime.

We have shown long range phase coherence of fermion pairs in an
optical lattice in the BEC-BCS crossover region by observing sharp
interference peaks during ballistic expansion. This indicates that
we have achieved strong $s$-wave pairing and superfluidity in a
lattice potential.  Further studies will reveal how the pair
wavefunction is affected by confinement \cite{mori05conf}, and
whether the lattice shifts the BEC-BCS crossover away from the
Feshbach resonance \cite{stoof06}. The loss of coherence during
the lattice ramp up and the rapid recoherence are characteristic
of a Mott-insulator.  However, definitive proof requires a better
understanding of the unitarity-limited interactions in such a
Fermi system. Recent theoretical work \cite{duan05, gubb06mol}
predicts that strongly interacting fermions in an optical lattice
feature multi-band couplings and next-neighbor interactions and
can realize the important $t-J$ and magnetic $XXZ$ models of
condensed matter theory.  This demonstrates that such atomic
systems are an ideal laboratory for the exploration of novel
condensed-matter physics.

\section*{Methods}

Clouds of superfluid fermion pairs were created in a new
experimental setup \cite{zoran03fifty,stan05oven} using techniques
similar to those described elsewhere \cite{zwie05Vort}. In brief,
a combination of laser cooling and sympathetic cooling of spin
polarized fermions by bosonic \na\ was followed by a spin transfer
to create a two component Fermi gas, allowing further cooling via
direct evaporation of the fermions. As the fermions cooled, they
formed pairs which Bose-condensed.

Estimates of the scattering length and hence the interaction
parameter from magnetic field were obtained using
$a(B)=-1405a_0(1+300/(B-834))(1+0.0004(B-834))$ \cite{bart04fesh},
where $B$ is measured in Gauss, and $a_0$ is the Bohr radius. The
calibration of the magnetic field in our system had an uncertainty
of about 5 G.

Evaporation was performed at a magnetic field of 822 G, where
strong interactions allowed for efficient evaporation. An
estimated average final number of $N\simeq 2\times 10^5$ \li\
pairs and harmonic trapping frequencies of $\nu_{x,y,z} =
(270,340,200)$ Hz gave a trap depth of 1.7 $\mu$K and Fermi energy
$E_F = k_B\times 1.4$ $\mu$K, where
$E_F=\hbar\overline{\omega}(6N)^{1/3}$ and $\overline{\omega}$ is
the average trapping frequency. After evaporation, the magnetic
field was brought to a desired value $B_0$ in 20 ms and the
condensate allowed to equilibrate for a further 200 ms. Before
ramping to values of $B_0$ on the BCS side, we also recompressed
the optical trap to $(340,440,270)$ Hz and 2.2 $\mu$K depth in 100
ms to accommodate the larger Fermi clouds above the resonance
\cite{zwie04rescond}.

A three-dimensional optical lattice was formed from three optical
standing waves, oriented such that the resulting unit cell has a
sheared cubic structure, with one axis tilted $\sim$ 20 degrees
from the normal for reasons of optical access (see Fig. 1a)
\cite{xu05}. The incident laser beams were focused down to the
condensate with waists of $\sim$ 90 $\mu$m, then retro-reflected
and overlapped at the condensate to generate the standing wave
potentials. All lattice light was derived from a 1064 nm single
frequency fiber laser, and each beam was detuned tens of MHz with
respect to the others to eliminate interference between different
beams.

The lattice potential was imposed onto the condensate by
adiabatically increasing the intensity of the laser beams to a
variable final value $V_0$.  The calibration of $V_0$ had an
uncertainty of about 20 percent. A simple linear ramp with a
constant rate $dV_0/dt$ of 0.5 \Er\ per ms was used unless
otherwise specified. This satisfies the interband adiabaticity
condition of $dV_0/dt<<16E_r^2/\hbar$.

Ballistic expansion for the detection of the different momentum
components was provided by a magnetic field sequence (shown in
Fig. 1) that quickly brought the system out of the strongly
interacting regime when all confinement was switched off. During
the magnetic field ramp of about 150 $\mu$s, the lattice potential
was kept on. The first 2 ms of expansion took place at 470 G,
where the molecules are tightly bound, before the field was ramped
back up to 730 G in the next 4.5 ms, at which the weakly
bound molecules strongly absorb light near the atomic resonance
line and could be observed by absorption imaging. The specific
magnetic field sequence was chosen to minimize collisions within
technical capabilities.

We would like to thank E. Demler, Z. Hadzibabic and M. Zwierlein
for helpful discussions. This work was supported by the NSF, ONR,
and NASA.

\end{document}